\documentclass[conference]{IEEEtran}

\usepackage{cite}

\usepackage{graphicx}

\begin{document}
%
\title{Design of Dual-Polarization Horn-Coupled Kinetic Inductance Detectors for Cosmic Microwave Background Polarimetry}

\author{\IEEEauthorblockN{Sean Bryan, George Che,\\and Philip Mauskopf}
\IEEEauthorblockA{Arizona State University\\
Tempe, AZ, USA\\
Email: sean.a.bryan@asu.edu}
\and
\IEEEauthorblockN{Kristi Bradford, Daniel Flanigan,\\ Bradley R. Johnson, Glenn Jones,\\Bjorn Kjellstrand, Michele Limon,\\Heather McCarrick, Amber Miller,\\and Brian Smiley}
\IEEEauthorblockA{Columbia University\\
New York, NY, USA}
\and
\IEEEauthorblockN{Peter Day}
\IEEEauthorblockA{Jet Propulsion Laboratory\\
Pasadena, CA, USA}}

\maketitle

\begin{abstract}
Mapping the polarization of the Cosmic Microwave Background is yielding exciting data on the origin of the universe, the reionization of the universe, and the growth of cosmic structure. Kilopixel arrays represent the current state of the art, but advances in detector technology are needed to enable the larger detector arrays needed for future measurements. Here we present a design for single-band dual-polarization Kinetic Inductance Detectors (KIDs) at 20$\%$ bandwidths centered at 145, 220, and 280 GHz. The detection and readout system is nearly identical to the successful photon-noise-limited aluminum Lumped-Element KIDs that have been recently built and tested by some of the authors. Fabricating large focal plane arrays of the feed horns and quarter-wave backshorts requires only conventional precision machining. Since the detectors and readout lines consist only of a single patterned aluminum layer on a SOI wafer, arrays of the detectors can be built commercially or at a standard university cleanroom.
\end{abstract}

\IEEEpeerreviewmaketitle

\section{Introduction}

Measurements of the polarization anisotropies of the Cosmic Microwave Background (CMB) radiation are revealing information about cosmic inflation~\cite{baumann09}, re-ionization~\cite{zaldarriaga08}, and the large-scale structure of the universe \cite{smith08}. Since current instruments are photon-noise limited, and already have dedicated observatories with months or years of integrating time available, moving forward to the next generation of CMB measurements will require larger detector arrays of devices that still maintain the high optical performance of currently deployed instruments.

The highly multiplexed readout and ability to fabricate large numbers of detectors on a single wafer make kinetic inductance detectors (KIDs) natural candidates for large detector arrays. Among other KID-based instruments, NIKA \cite{catalano14} has already successfully observed and published science results at 140 GHz and 240 GHz, which are frequencies ideal for CMB measurements. Also, in lab testing some of the authors have demonstrated single-polarization KIDs, with high optical efficiency and photon-noise limited performance down to low optical loading levels \cite{mccarrick14}. These successes motivate dual-polarization KID detector designs for future CMB instruments. The ideal pixel in a detector array for this application would have a single feed with detectors each sensing one of multiple passbands, and would also have another identical set sensing the other polarization. In this paper, we consider a more straightforward single-band dual-polarization design.

\section{Design}

This detector architecture is designed to enable large arrays of sensitive single-band detectors that can be fabricated commercially or at facilities widely available at universities. The feed horns for an entire array module will be machined on a single metal part using direct drilling. A second metal part will form the array of backshorts and hold the detector wafer for the array. It will be precisely registered to the horn array using standard alignment pins. Finally, the detector wafer will also contain the readout lines, and the KID architecture allows a large number of detectors to be read out with a single RF coax and low-noise amplifier.

\begin{figure*}
\begin{center}
\includegraphics[width=0.29\textwidth]{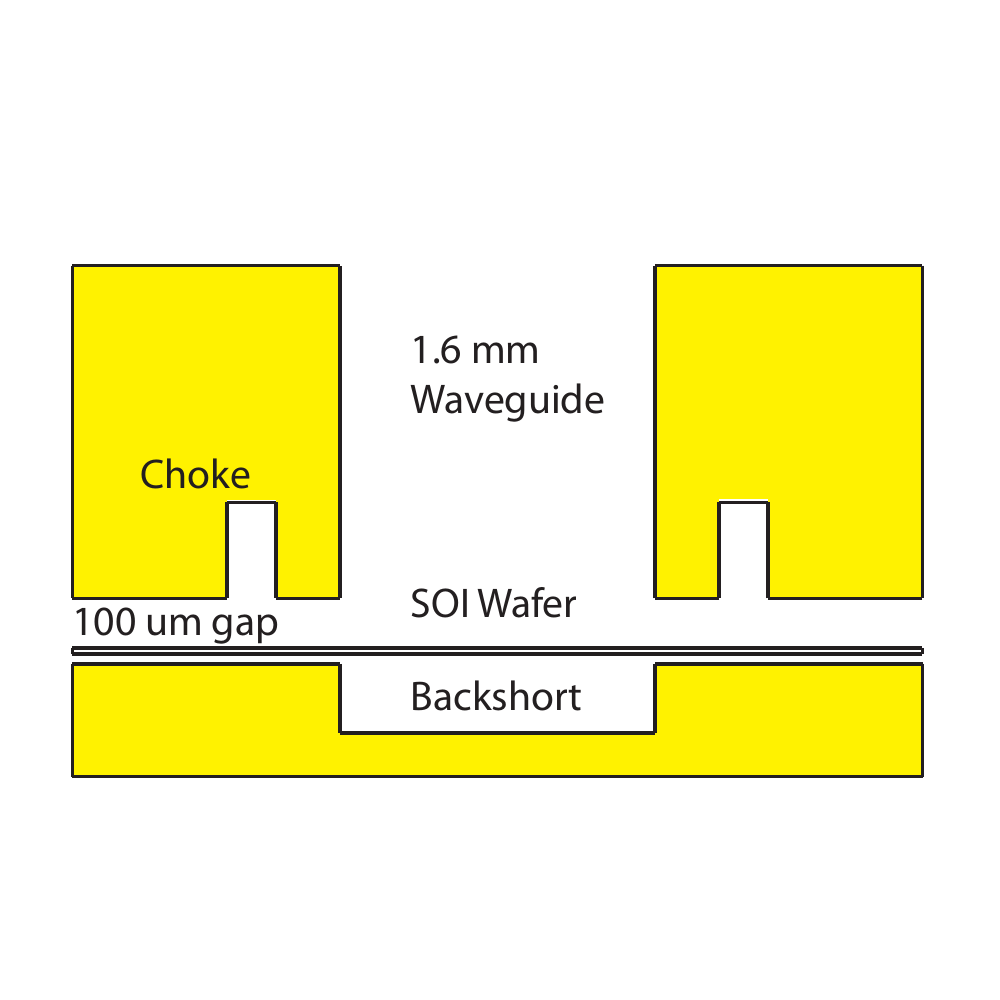}
\includegraphics[width=0.372\textwidth]{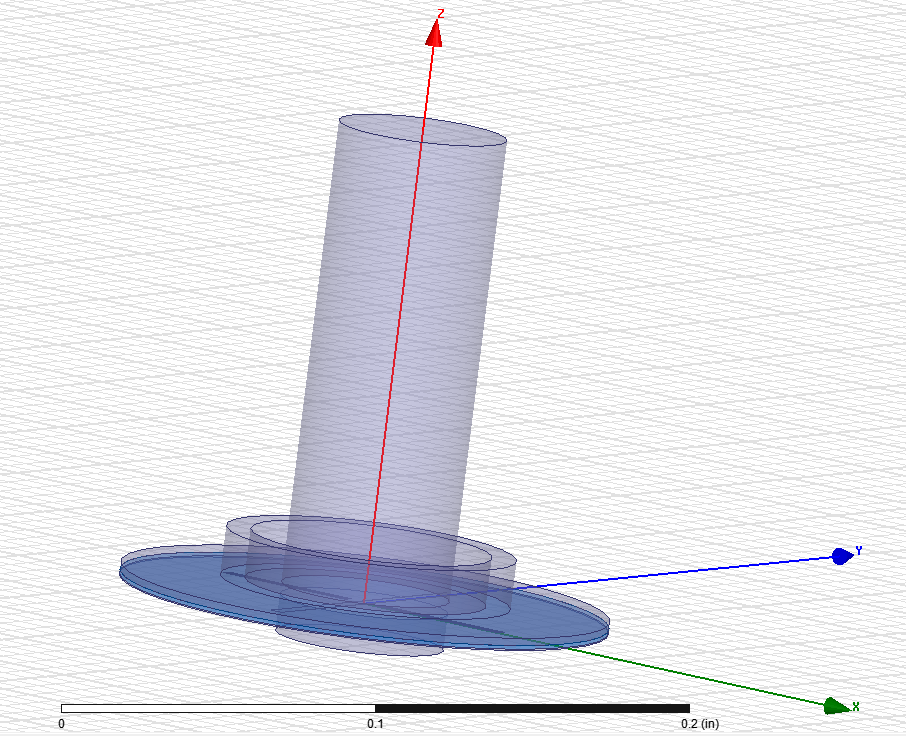}
\includegraphics[width=0.32\textwidth]{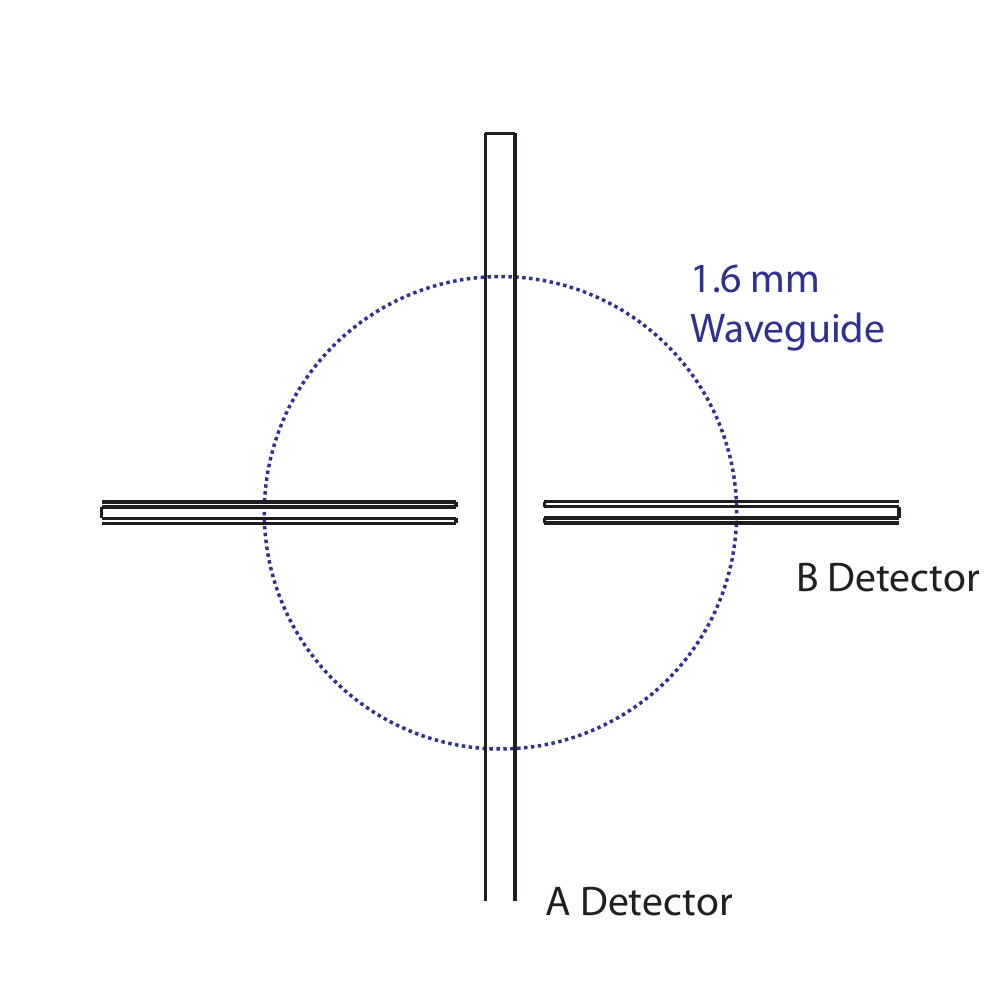}
\caption{Drawings of the 145 GHz pixel. A 3D view of the HFSS drawing is shown in the center, and a cross-section (not to scale) is shown in the left. The top is a waveguide section (a feed horn will be above, and is not shown in this simulation). The silicon layer of the SOI wafer is below it, through which the aluminum LEKIDs are back-illuminated. A quarter-wave backshort is below. A 100 micron air gap allows the detector wafer to be mounted between the top and bottom metal structures, and a waveguide choke is used to prevent light from leaking through this gap. The right shows a top view of the absorber/inductor portion of the LEKID detector pair for a single device. \label{design}}
\end{center}
\end{figure*}

A smooth walled circular feed horn will couple the incident light onto a circular waveguide and then onto the detectors. Smooth walled circular feed horns can be designed with a profile that launches a circularly-symmetric tapered beam with low cross-polarization. They are also straightforward to fabricate using direct drilling with a custom tool made to create the profile. Measurements at 700 GHz of horns made with this approach show good performance \cite{tan12}. The cutoff of the circular waveguide section below the horn will define the lower edge of the detector's passband. The upper edge will be defined by installing a lowpass filter directly above the detector array \cite{ade06}.

The light will be coupled onto aluminum lumped-element KIDs (LEKIDs) \cite{mccarrick14}. The detectors demonstrated in the paper were fabricated commercially, since they consist only of a single layer of aluminum deposited on a substrate. The lumped elements used to form the readout resonator allow the readout frequencies of the detectors to lie in the RF, around 1 GHz, which enables readout using standard microwave electronics. For the purposes of this abstract, we are only modeling the inductor/absorber portion of the detector. We will subsequently modify the existing capacitor and readout architecture.

The detectors are fabricated on a chip. We will use a silicon-on-insulator (SOI) wafer as the chip, which will allow us to put the aluminum absorbers on the thin silicon layer and etch away the rest of the wafer behind the optically-active region. We will leave the rest of the wafer at its full handle thickness, enabling a robust mechanical mount of the wafer in the horn assembly. Even with the thin silicon, there will be an air gap between the horn array and the backshort array. That airgap would let radiation leak out, causing a loss in optical efficiency and possibly inducing optical cross talk between devices in the detector array. To prevent this, we will fabricate a circular groove around each waveguide, forming a choke joint. A standard precision end mill can machine this feature. A cross section view showing the waveguide, backshort, and choke is shown in Figure~\ref{design}. The electrical length from the choke to the waveguide is such that it reflects most of the leaking radiation back into the detector cavity for absorption.

\begin{figure}
\begin{center}
\includegraphics[width=0.4\textwidth]{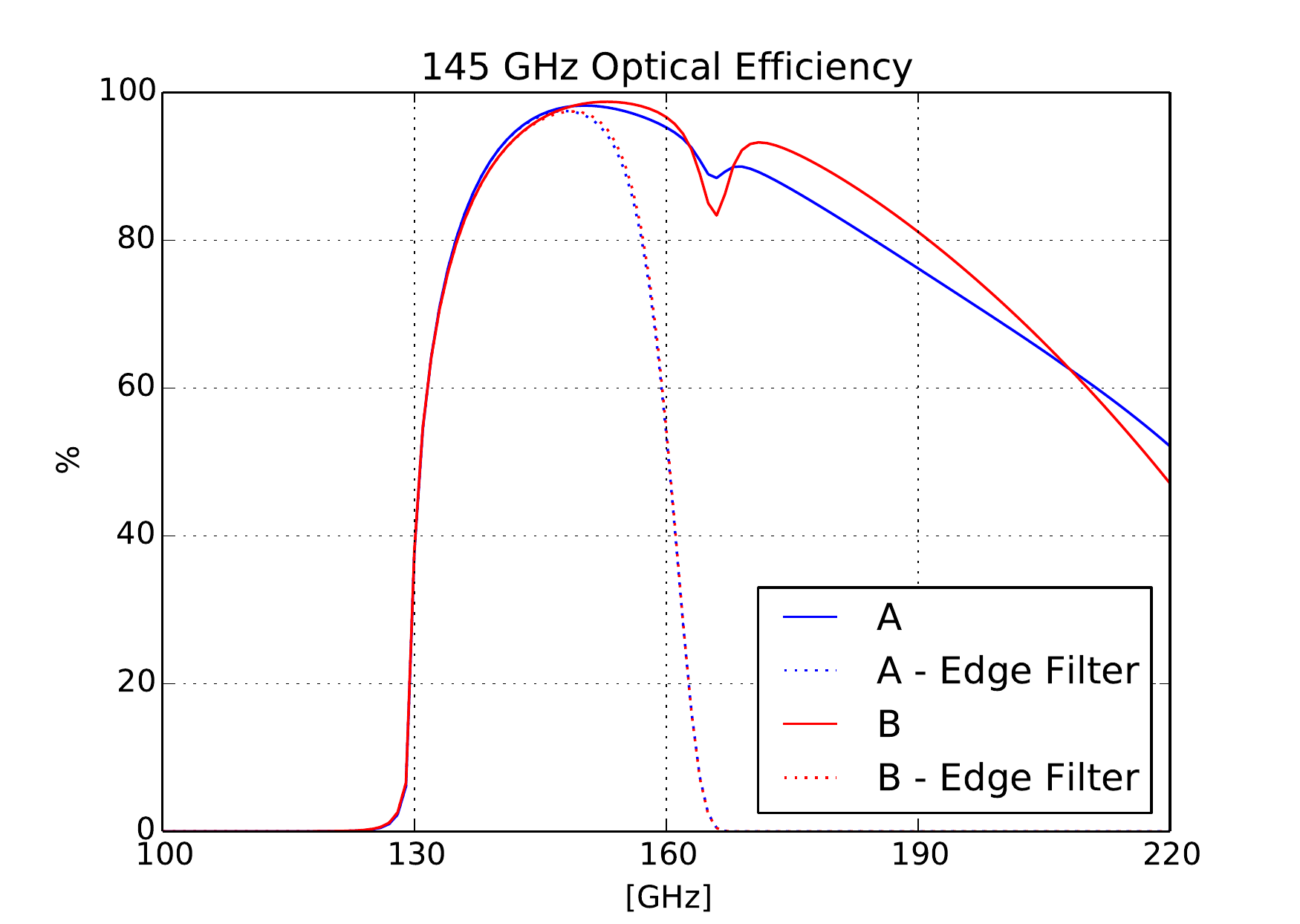}
\caption{Simulated optical efficiency of the 145 GHz design. The solid lines show the power absorbed in the A- and B-polarized detectors. The waveguide section prevents light below the cutoff frequency from being absorbed, defining the lower band edge. The dotted lines show the absorption after multiplying by a nominal edge filter which will be installed to define the upper band edge. \label{l150}}
\end{center}
\end{figure}

We simulated the optical efficiency from the base of the horn through to the absorption onto the detector using HFSS. The layout of the inductor/absorbers is shown in Figure~\ref{design}. They are 2 micron linewidth aluminum traces, which we model as having a surface impedance of 4 ohms per square informed by earlier measurements of the single-polarization devices. After fixing the dimensions of the absorbers, the airgap, and the waveguide diameter, we used the Matlab API for HFSS to calculate the optical efficiency across the passband for many design parameter values. The parameter sweep was over the choke location, choke width, choke depth, and the depth of the backshort. Each individual simulation takes several minutes, so it takes days of workstation time to generate the hundreds of simulations needed to explore the parameter space. The simulated optical efficiency of both detectors for a 20$\%$ bandwidth device centered at 145 GHz is shown in Figure~\ref{l150}.

\section{Scaling the 145 GHz design to higher frequencies}

\begin{figure}
\begin{center}
\includegraphics[width=0.50\textwidth]{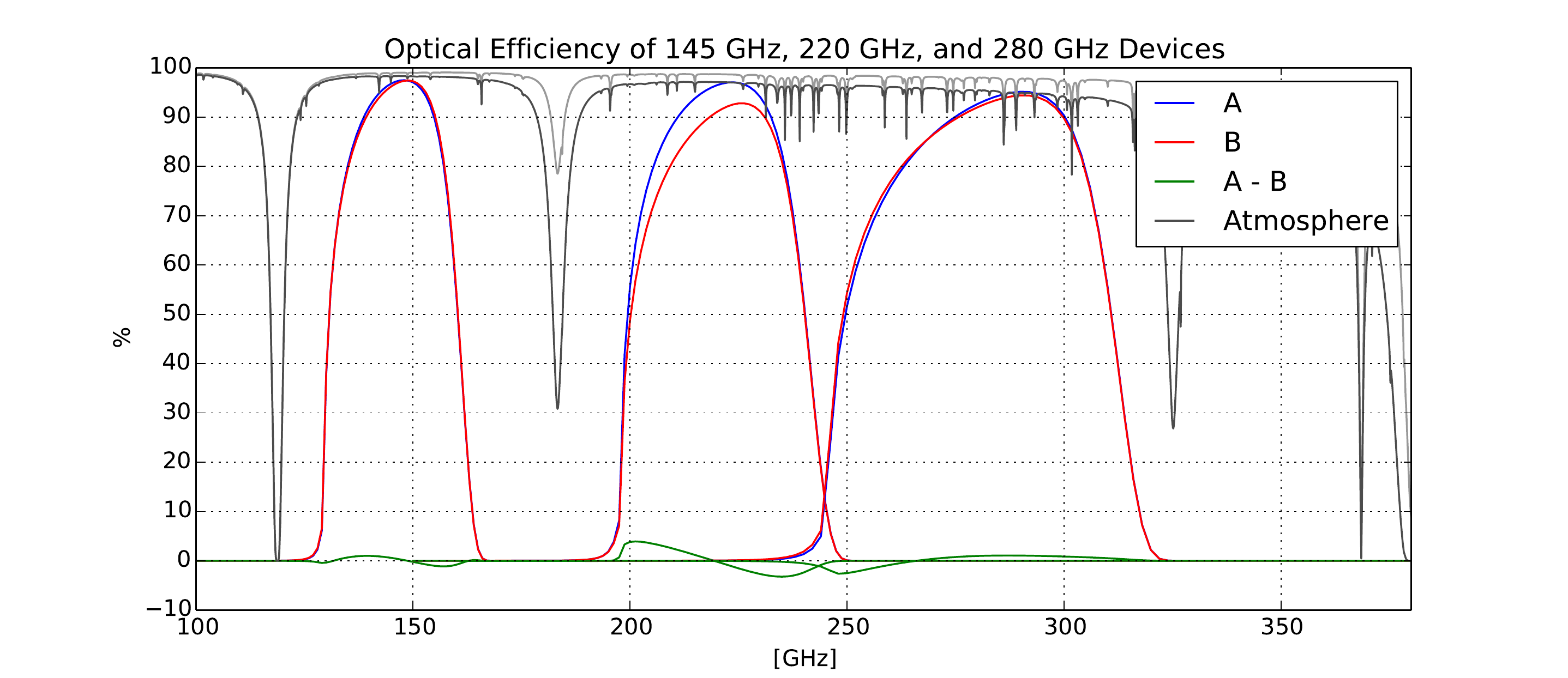}
\includegraphics[width=0.50\textwidth]{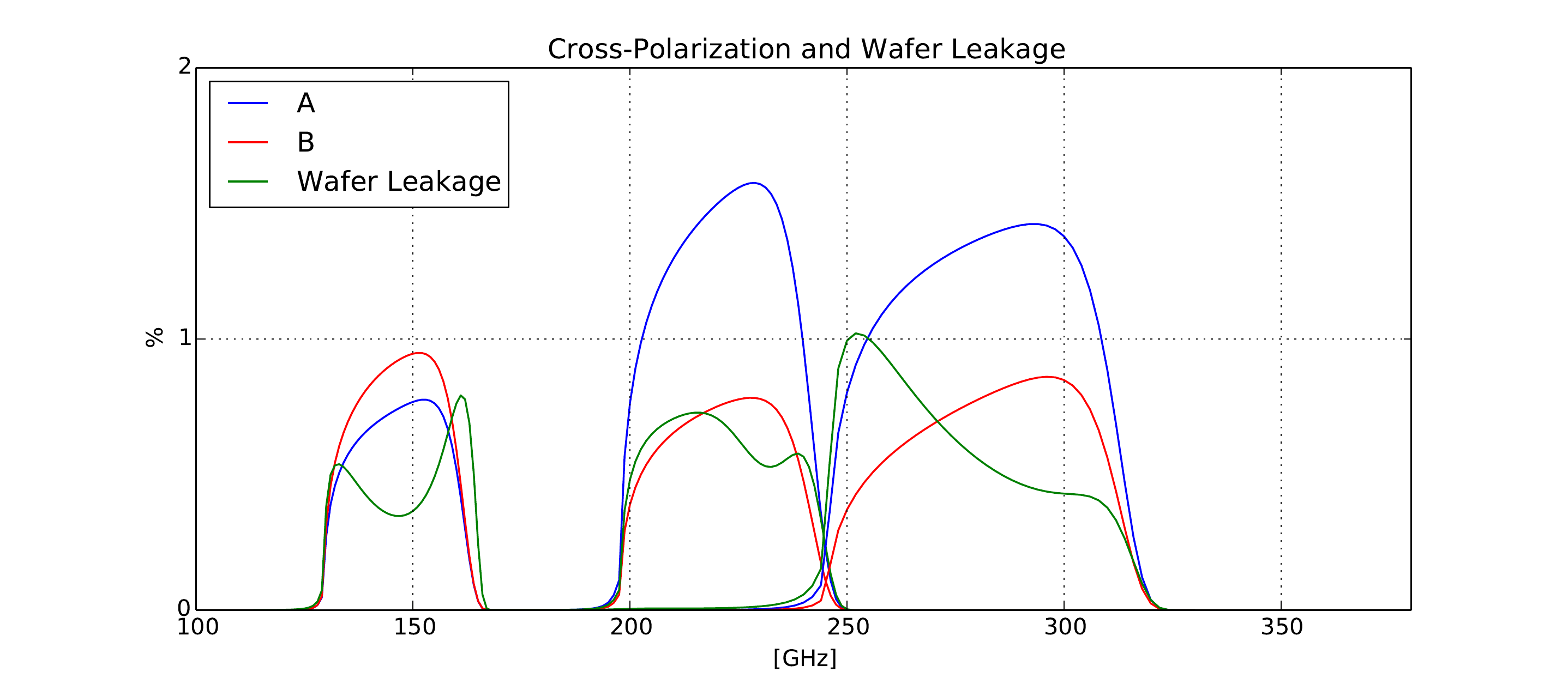}
\caption{Performance of the three single-band detector designs. The top panel shows the passbands of the A and B detectors of each of the three devices, and the AB spectral mismatch, corrected for the slight gain difference between the two devices, is shown in green. A model of atmospheric transmission (from the ALMA sky model) is also shown. The bands avoid the strong atmospheric absorption lines near 120 GHz, 180 GHz, and 320 GHz. The bottom panel shows the cross-polarization, which is below 2 percent for every device. The light leaking through the airgap onto all four (or six for a hex-packed array) nearest-neighbor detectors is shown in green, so the optical cross-talk between a single pair of devices would be a factor of 4 or 6 smaller than the line shown. \label{threeband}}
\end{center}
\end{figure}

The CMB anisotropy signal peaks near 150 GHz, and our design for a 145 GHz device looks promising. However, the desire to remove foreground contamination motivates observations at nearby frequency bands as well. Since the detection mechanism of KIDs is breaking Cooper pairs, a 90 GHz aluminum LEKID device would be marginal because the photons are only marginally energetic enough to be detected. Going to lower frequencies would require switching to a superconductor with a lower transition temperature than aluminum, and therefore more weakly bound Cooper pairs. However, higher frequency designs are possible with aluminum, which would enable this technology to be used for observations of galactic dust that is thought to be a dominant foreground for CMB polarization measurements. Higher frequency observations are also used for studies of secondary anisotropies like the SZ effect.

We scaled the 145 GHz design up to 280 GHz. Because the linewidth of the aluminum absorbers cannot be scaled smaller than the 2 microns used in the 145 GHz design, and because we switched to a 10 micron silicon layer thickness for the 280 GHz design, we performed a second parameter sweep to optimize the 280 GHz design. We also scaled the design to 220 GHz, with a 20 micron silicon layer, and performed another optimization sweep there. The spectral performance for all three single-band devices is shown in Figure~\ref{threeband}.

The optical efficiency of the devices at all three bands is calculated to be very high, which motivates finalizing the wafer design and moving towards optical testing. The spectral matching between the A and B detectors is good, within $\pm$1$\%$ across the entire band in the 145 GHz, $\pm$4$\%$ for the 220 GHz device, and $\pm$2$\%$ for the 280 GHz device. The cross polarization is below 2$\%$ in all devices. The total light leaking through the air gap is roughly 1$\%$. Since this leakage is going onto all four nearest neighbor pixels, or six in the case of a hex-packed array, the approximate optical cross talk between a pair of pixels is a factor of 4 or 6 smaller, i.e. smaller than 1$\%$.

\section{Conclusions}

We presented a design for a single-band dual-polarization detector that could be used in large detector arrays for CMB measurements. The device is simulated to have high optical efficiency, good spectral matching between the two polarized detectors, and low cross polarization and optical cross talk. The devices could be fabricated commercially, with the wafers made at a device foundry and the horns and backshorts made at a precision machine shop. Also, many universities have the cleanroom facilities and precision milling machines needed for the fabrication. Future work will include finalizing the design, choosing a feed horn design, and fabrication and initial testing of a small array of devices.

\bibliographystyle{unsrt}
\bibliography{bibliography}

%
%

\end{document}